\documentclass[aps,preprint,prx,longbibliography]{revtex4-1}
\pdfoutput=1
\usepackage[utf8]{inputenc}
\usepackage{amsmath,amssymb,mathrsfs,amsthm}
\usepackage{bbm}
\usepackage{chemarrow}
\usepackage{graphicx}
\usepackage{accents}
\usepackage{sidecap}
\usepackage[colorlinks=true,allcolors=blue]{hyperref}
\usepackage{breakurl}
\usepackage{soul}
\usepackage{float}
\usepackage[shortlabels]{enumitem}
\usepackage{epstopdf}

\def\rd{{\rm d}}
\def\va{\boldsymbol{a}}
\def\vb{{\bf b}}

\def\vp{{\bf p}}

\def\vx{{\bf x}}
\def\vy{{\bf y}}
\def\vu{{\bf u}}

\def\vnu{\boldsymbol{\nu}}

\def\vxi{\boldsymbol{\xi}}

\def\mT{{\bf T}}
\def\vX{{\bf X}}
\def\vY{{\bf Y}}
\def\vZ{{\bf Z}}

\begin{document}

\title{On Thermodynamic Information}

\author{Bing Miao}
\affiliation{Center of Materials Science and Optoelectronics Engineering, College of Materials Science and Opto-Electronic Technology, University of Chinese Academy of Sciences, Beijing 100049, P.R.C.}

\author{Hong Qian}
\affiliation{Department of Applied Mathematics, University of Washington, Seattle, WA 98195, U.S.A.}

\author{Yong-Shi Wu}
\affiliation{Department of Physics \& Astronomy, University of Utah,\\ Salt Lake City, UT 84112, U.S.A.}


\begin{abstract}
Information based thermodynamic logic is revisited.  It consists of two parts: Part A applies the modern theory of probability in which an arbitrary convex function $\phi$ is employed as an analytic ``device'' to express information as statistical dependency contained in the topological sub-$\sigma$-algebra structure.  Via {\em thermo-doubling}, Fenchel-Young equality (FYE) that consists of $\phi(\vx)$ and its conjugate $\psi(\vy)$ establishes the notion of {\em equilibrium} between $\vx$ and $\vy$ through duality symmetry and the principle of maximum entropy/minimum free energy.  Part B deals with a given set of repetitive measurements, where an inherent convex function emerges via the mathematics of large deviations. Logarithm-based Shannon entropy with $\phi(x)=-\log x$ figures prominently for i.i.d. sample statistics.  Information can be a measure of the agreement between a statistical observation and its theoretical models. Maximum likelihood principle arises here and FYE provides a thermodynamic energetic narrative of recurrent data.
\end{abstract}

\maketitle

\newpage
\tableofcontents


\section{Introduction}

The current understanding of {\em information} as a scientific subject and its quantification, outside quantum physics, seems to be chiefly defined by the Shannon entropy and its variations such as Kullbeck-Leibler divergence \cite{cover-book,hobson_1969}, although a sizable literature on nonextensive entropy exists \cite{karmeshu-book,hongliu_tsallis}.  On the other hand, in the teaching of modern mathematical theory of probability \cite{KolmogorovBook,DurrettBook}, the English word ``information'' is routinely used to refer the rich structure within the $\sigma$-algebra that is at the foundation of a probability measure, and the statistical concept of ``conditioning''.  The present work revisits the theory of information and its relation to thermodynamics: We show the notion of information is best understood through a two-part theory from at least three different perspectives: As an integral part of Kolmogorov's $(\Omega,\mathcal{F},\mathbb{P})$, information that is contained in $\mathcal{F}$ and its various sub-$\sigma$-algebra $\mathcal{G}\subset\mathcal{F}$ \cite{urbanik} can be analytically brought out as a collection of inequalities via an arbitrary convex function.  All these results are consequences of the well-known Jensen's inequality applied to the conditional expectation of a random variable $\vX$,
\begin{equation}        \phi\Big(\mathbb{E}\big[\vX\big|\mathcal{G}\big]\Big)
    \le \mathbb{E}\big[ \phi(\vX)\big|\mathcal{G} \big],
\label{eq0}
\end{equation}
where $\phi$: $\mathbb{R}\to\mathbb{R}$ is a convex function.  On the left of (\ref{eq0}) is a partially (over $\mathcal{G}$) pre-averaged $\vX$, thus its $\phi$-based ``information'' decreases when compared with the average of $\phi(\vX)$ over $\mathcal{G}$.  We shall extend this idea to Markov dynamics and show a host of Boltzmann's $H$-theorem like inequalities can be obtained, that include entropy production and free energy dissipation.  The logarithmic convex function and the Shannon entropy, however, are particularly important for certain additivity of trajectory-based stochastic entropy production.

The relation between above $\phi$-information inequality and thermodynamics is based on Legendre-Fenchel transform (LFT) \cite{lu-qian-22,qian_jctc}:  Corresponding to any $\phi(\vx)$, $\vx\in\mathbb{R}^K$ there is a convex \cite{lu-qian-22}
\[
    \psi(\vy) = \sup_{\vx\in\mathbb{R}^K}\big\{
      \vx\cdot\vy - \phi(\vx) \big\},
\]
and together they provide the Fenchel-Young inequality under a thermodynamic state space doubling $ (\vx,\vy)\in\mathbb{R}^K\otimes\mathbb{R}^K$, {\em thermo-doubling} for short:
\[
    \eta(\vx,\vy):=\phi(\vx)+\psi(\vy)- \vx\cdot\vy \ge 0.
\]
We identify $\eta$ as the {\em entropy production}  {\em \`{a} la} the Brussels school of nonequilibrium thermodynamics \cite{prigogine-book}, and $\eta\ge 0$ as the Second Law \cite{lieb}.  Equilibrium is between $\vx$ and its corresponding conjugate variable $\vy^{\text{eq}}=\nabla\phi(\vx)$, or equivalently $\vx^{\text{eq}}=\nabla\psi(\vy)$. $\eta(\vx,\vy)=0$ defines a $K$-dimensional equilibrium manifold within the $(K+K)$-dimensional, doubled thermodynamics state space. Equilibrium possesses duality symmetry \cite{lu-qian-22} and implies a pair of variational principles:
\begin{subequations}
\begin{eqnarray}
      \vx^{\text{eq}}(\vy) &=& \arg\sup_{\vx\in\mathbb{R}^K}\big\{\vx\cdot\vy - \phi(\vx) \big\}  \,  \text{ for a given } \vy \text{ and varying } \vx;
\\
      \vy^{\text{eq}}(\vx) &=& \arg\inf_{\vy\in\mathbb{R}^K}\big\{\vx\cdot\vy - \psi(\vy) \big\}  \,  \text{ for a given } \vx \text{ and varying } \vy.
\end{eqnarray}
\end{subequations}

The ``information'' can be brought out by any convex function $\phi$ in above Part A. Numerical value aside, it expresses the topological characteristics of $\sigma$-algebra in connection to statistical dependency and nonlinear correlations.  With thermo-doubling, LFT furnishes a thermodynamic, ``energetic'' narrative with nonequilibrium entropy production $\eta>0$ and equilibrium  symmetry and conservation with $\eta=0$.  Then in Part B, when a specific set of repeated measurements with data {\em ad infinitum} is explicitly considered, a particularly relevant convex function emerges from the theory of large deviations \cite{qian_jctc,yangqian_22,dembo-book}.  In this case, large deviation rate function is a bivariate convex function $\phi(\vx\|\vxi)$ which quantifies the amount of information in the data $\vx$ w.r.t. to a probabilistic model or model parameters $\vxi$. Viewed differently, $\phi(\vx\|\vxi)$ measures the goodness of a statistical model captured by $\vxi$ w.r.t. the empirically observed $\vx$ \cite{qian_jctc,paper-I}.

The present work fulfills E. T. Jaynes' vision \cite{JaynesBook} of unifying information theory and  statistical thermodynamics and bridging Kolmogorov's mathematical theory with statistical data modeling. The paper is arranged as follows:
Sec. \ref{sec:2part} presents the two-part theory, in \ref{sec:2.A} and \ref{sec:2.B} respectively.  Sec. \ref{sec:2.A} also contains the basic material on conditional expectation for readers who are not familiar with the mathematics.  A key issue to note is that $\mathbb{E}[\vX|\mathcal{G}]$ is still a function of $\Omega\to\mathbb{R}$, just as $\mathbb{E}[\vX]$ being a ``trivial'' function with each and every $\omega\in\Omega$ taking the same value: Though extremely limited, $\mathbb{E}[\vX]$ nevertheless still provides some information on $\vX$!  Sec. \ref{sec:2.C} then shows that, for finite $\Omega$ with $\|\Omega\|=K$, conditional expectation can be understood as a non-invertible linear transformation $\mT_{\mathcal{G}}\vX$ for $\vX\in\mathbb{R}^K$, the space of all possible random variables as $K$-vectors.

Sec. \ref{sec:markov} studies Markov dynamics with discrete time and finite state space.  $\phi$-based generalization of the notions of free energy dissipation and entropy production from current stochastic thermodynamics \cite{peliti-book,shiraishi-book} is investigated.  Sec. \ref{sec:cg} continues the discussion on non-invertible linear transformations on $\mathbb{R}^K$, as coarse-graining and/or measurements with incomplete information, and related LFTs, maximum entropy principle, and effective thermodynamics.  The paper concludes with discussions in Sec. \ref{sec:5}.

\section{Information, Stochastic Entropy, and Thermodynamics}
\label{sec:2part}

The idea that entropy is itself fluctuating as a random variable $\Upsilon$ can be traced back to the Radon-Nikodym derivative (RND) formulation of trajectory-based entropy production in Markov dynamics \cite{qian_pre_eec,qqg-contemp}, and even earlier work of Kolmogorov \cite{kolmogorov-information} and Tribus \cite{tribus-book}.  More precisely: $\Upsilon:=-\log f_Y(Y(\omega))$, $\omega\in\Omega$, where $f_Y(y)$, $y\in\mathbb{R}$ is the probability density function of a random variable $Y$. It immediately follows that corresponding to an observed statistical change $f_1(y)\to f_2(y)$, one should consider
\begin{equation}
    \Delta\Upsilon = -\log\left(\frac{f_2(Y)}{f_1(Y)}\right) = -\log\left(\frac{\rd\mathbb{P}_2}{\rd\mathbb{P}_1}(\omega)\right).
\end{equation}
The second equality holds if the random variable $Y$, as an observable, is $\mathcal{F}$-measurable.  In other words, $Y$ resolves all the information in $(\Omega,\mathcal{F})$.   One then immediately has an equality and an inequality:
\begin{equation}
     \mathbb{E}^{\mathbb{P}_1}\Big[ e^{-\Delta\Upsilon} \Big] = 1 \, \text{ and } \, \mathbb{E}^{\mathbb{P}_1} \big[\Delta\Upsilon\big] \ge 0.
\end{equation}
They are the mathematical incarnations of the Jarzynski equality and the Clausius inequality \cite{jarzynski}.  More interestingly, if $Y$ provides only partial information, i.e., $\mathcal{G}:=\sigma(Y)$ is only a sub-$\sigma$-algebra $\mathcal{G}\subset\mathcal{F}$, then
\begin{equation}
\label{info_ineq}
  \mathbb{E}^{\mathbb{P}_1}\left[ -\log\left(\frac{f_2(Y)}{f_1(Y)}\right) \right]  \le\mathbb{E}^{\mathbb{P}_1}\left[-\log\left(\frac{\rd\mathbb{P}_2}{\rd\mathbb{P}_1}(\omega)\right)\right].
\end{equation}
This is known as information inequality \cite{cover-book}.

\subsection{Information inequality}
\label{sec:2.A}

To provide a deeper understanding of Eq. (\ref{info_ineq}), we focus on probability space $(\Omega,\mathcal{F},\mathbb{P})$ with finite $\Omega=\{1,\cdots,K\}$, $\mathcal{F}=2^{\Omega}$, and a random variable $\vX=(x_1,\cdots,x_K)\in\mathbb{R}^K$, whose expectation
\[
 \mathbb{E}[\vX]=\sum_{k=1}^K p_kx_k.
\]
Consider a sub-$\sigma$-algebra $\mathcal{G}$ that represents a partition of $\Omega$:
\begin{equation}
   \Omega = \bigcup_{g=1}^G
    \tilde{\Omega}_{g}, \text{ where }
    \, \tilde{\Omega}_i\bigcap\tilde{\Omega}_j = \emptyset \, \text{ when } \,
        i\neq j,
\end{equation}
$\|\tilde{\Omega}_g\|=K_g$, and
$K_1+\cdots+K_G=K$.   We shall re-lable the elements in $\Omega$ as $(i,j)$, where
$(i, j)\in\tilde{\Omega}_i$ for $1\le j\le K_i$.  Then the conditional expectation is
\begin{equation}
    \mathbb{E}[\vX|\mathcal{G}]
    = \frac{1}{\mathbb{P}\big(\tilde{\Omega}_i\big)}
    \sum_{k=1}^{K_i} p_{ik}x_{ik}, \
    \mathbb{P}\big(\tilde{\Omega}_i\big)
      = \sum_{k=1}^{K_i} p_{ik},
\end{equation}
for state $(i,j)\in\Omega$. $\mathbb{E}[\vX|\mathcal{G}]$: $\Omega\to\mathbb{R}$ is a random variable with limited information: It cannot differentiate the different states within each $\tilde{\Omega}$. In physics, one adapts the viewpoint that $\mathbb{E}[\vX|\mathcal{G}]$ defines a smaller state space; but the theory of probability articulates a smaller $\sigma$-algebra on the same $\Omega$.

The profoundness of the mathematical concept of conditional expectation is as follows: First, we need to firmly assert that, for continuous $\Omega$, the probability $\mathbb{P}$ in $(\Omega,\mathcal{F},\mathbb{P})$ is not defined for elementary events,  $\omega\in\Omega$; rather it is defined on subsets,  $A\subset\Omega$, where $A\in\mathcal{F}$.  In plain words: One does not have a probability for a point $\omega$, it is zero.  One assigns a probability to a patch of points $A$.  A continuous random variable $\vX(\omega)$ as a scientific observable is a function of $\omega\in\Omega$.  The conditional expectation $\mathbb{E}[\vX|\mathcal{G}]$, $\mathcal{G}\subset\mathcal{F}$, is still a random variable; it is again a function of $\omega\in\Omega$.  The smallest elements in $\mathcal{G}$ are not all $\omega$, but already subsets of $\Omega$; just like lines and planes in $\mathbb{R}^3$.  The values of $\mathbb{E}[\vX|\mathcal{G}]$ are from a pre-averaging according to the probability $\mathbb{P}$. Therefore, if one believes that $\mathbb{P}$ is objective and intrinsic to a physical system,  then $\mathbb{E}[\vX|\mathcal{G}]$ provides incomplete information.  If, however, one believes that $\mathbb{P}$ is only a mathematical model of the system to be determined, then $\mathcal{G}$ contains all the remaining uncertainty and $\mathbb{E}[\vX|\mathcal{G}]$ is an average according to a suppositional model.

Suppose $\phi$: $\mathbb{R}\to\mathbb{R}$ is a twice differentiable convex function. Conditional Jensen's inequality
in Eq. (\ref{eq0}) yields
\begin{equation}
\label{23}
      \mathbb{E}\Big[\phi\Big(\mathbb{E}[\vX|\mathcal{G}] \Big) \Big]
        \le \mathbb{E}\Big[\mathbb{E}\big[\phi(\vX)\big| \mathcal{G} \big]\Big]
        = \mathbb{E}\big[\phi(\vX) \big],
\end{equation}
which is identified as the information inequality w.r.t. $\mathcal{G}$.  Moreover, corresponding to $\phi(x)$ is another twice differentiable convex function
\begin{equation}
    \psi(y)
    = \sup_{x\in\mathbb{R}} \big\{ xy
    - \phi(x) \big\},
\end{equation}
known as the LFT of $\phi$, and together they establish the Fenchel-Young inequality
\begin{equation}
        \eta(x,y)=\phi(x)+\psi(y) - xy \ge 0,  \
         x,y\in\mathbb{R},
\end{equation}
in which the equal sign holds true if and only if $y=\phi'(x)$ and simultaneously $x=\psi'(y)$ as a bijective relation between $x$ and $y$. One thus has
\begin{equation}
\label{fyie}
     \mathbb{E}\Big[
     \phi(\vX)+\psi(\vY) - \vX\vY
     \Big] \ge 0,
\end{equation}
where the equality holds true when $\vY=\phi'(\vX)$ and simultaneously
$\vX=\psi'(\vY)$, relations analogous to the Maxwell relations in equilibrium thermodynamics.  The equal sign in (\ref{fyie}) also echos the celebrated ``entropy $+$ free energy $-$ internal energy $=$ $0$'' in equilibrium thermodynamics.

Under a coarse-graining that is represented by
$\mathbb{E}[\vX|\mathcal{G}]$, one further has
\begin{equation}
      \mathbb{E}\Big[\phi\Big(
      \mathbb{E}[\vX|\mathcal{G}]\Big)
      + \psi(\vZ) - \vZ\,\mathbb{E}[\vX|\mathcal{G}] \Big] \ge 0,
\end{equation}
from which the ``equilibrium'' $\vZ^*$ corresponding to a given $\mathbb{E}[\vX|\mathcal{G}]$ is obtained via minimization,
\begin{eqnarray}
    \vZ^* &=& \arg\inf_{\vZ\in\mathcal{G}}
    \mathbb{E}\Big[ \phi\Big(
      \mathbb{E}[\vX|\mathcal{G}]\Big)
      + \psi(\vZ) - \vZ\,\mathbb{E}[\vX|\mathcal{G}] \, \Big]
\nonumber\\
      &=& \arg\sup_{\vZ\in\mathcal{G}}
    \mathbb{E}\Big[\vZ\,\mathbb{E}[\vX|\mathcal{G}]
      - \psi(\vZ)   \Big].
\end{eqnarray}
Actually,
\begin{equation}
    \psi'(\vZ^*) = \mathbb{E}[\vX|\mathcal{G}],
    \  \vZ^* = \phi'\big(\mathbb{E}[\vX|\mathcal{G}]\big),
\end{equation}
both are $\mathcal{G}$-measurable.  In general $\vZ^*\neq \mathbb{E}[\phi'(\vX)|\mathcal{G}]$;
the equation is valid only when $\phi'(x)$ is a linear function of $x$, which implies that $\phi(x)$ is quadratic.  This is the scenario of linear irreversibility and Gaussian fluctuations \cite{qian_prsa}.

The above results are mathematically true for any convex function $\phi$.  Eq. (\ref{info_ineq}) is merely a special case of (\ref{23}) with $\phi(x)=-\log x$, the random variable $\vX=\tfrac{\rd\mathbb{P}_2}{\rd\mathbb{P}_1}$, and
\[
  \mathbb{E}\left[ \left.    \frac{\rd\mathbb{P}_2} {\rd\mathbb{P}_1}
  \,\right|\mathcal{G} \right] =  \frac{f_2(Y)}{f_1(Y)}, \
  \mathcal{G}=\sigma(Y).
\]

\subsection{Data and information}
\label{sec:2.B}

With a given random variable $\vX$ as a scientific observable,  considering all the subsets in $\sigma(\vX)$ is anticipatory to all possible outcomes from a single measurement on $
\vX$.   Note that a measurement of $\vX$ is not necessarily a precise value $x$;  it could be any event in $\sigma(\vX)$: the collection of all subsets of $\Omega$ of the form $\{\omega: \vX(\omega)\in \mathcal{I}\}$ where $\mathcal{I}\in\mathcal{B}(\mathbb{R})$, the Borel $\sigma$-algebra of $\mathbb{R}$. The $\mathbb{E}[\phi(\vX)]$ in Sec. \ref{sec:2.A} is a quantification of the information content using the convex $\phi$ as an analytic ``device'', in the random variable $\vX$ w.r.t. $(\Omega,\mathcal{F},\mathbb{P})$.  From ``anticipatory'' to realization,  when an $A\in\sigma(\vX)$ is actually realized,  we say the amount of information is higher in an observation with smaller $\mathbb{P}(A)$.

In statistical analysis of recurrent data, among the arbitrary convex functions, there is a unique one that is defined by the nature of a statistical samples {\em ad infinitum}.  For independent and identically distributed (i.i.d.) samples, convex function $\phi(x)=-\log x$ is the unique one: In fact logarithm is the only function if one insists entropy additivity among statistically independent events \cite{shore-johnson}. With respect to a given measurement, logarithmic probability measures the amount of ``information'' in the data w.r.t. a probabilistic model; or from a different perspective it measures the ``goodness'' of the model w.r.t. the data.  This latter interpretation of information leads to
maximum likelihood principle.

Newtonian physics and modern theory of ergodic nonlinear dynamics \cite{qxz-book} revitalizes the frequentist's interpretation of probability. The unique convex function is the level II large deviations rate function in the theory of large deviations \cite{dembo-book}.  The maximum entropy principle then provides a host of entropy functions $\phi(\vx\|\vxi)$ as a bi-variate convex non-negative function of observations $\vx$ and its matching probabilistic model parameters $\vxi$, $\phi(\vxi\|\vxi)=0$ \cite{paper-I}.  Corresponding to $\phi(\vx\|\vxi)$ is its LFT $\psi(\vy,\vxi)$. This is the domain of thermodynamics \cite{lu-qian-22}: it provides an ``energetic narrative'' for the data $\vx$ from recurrent phenomena.  The Fenchel-Young inequality
\begin{equation}
    \eta(\vx,\vy) = \phi(\vx\|\vxi) +
    \psi(\vy,\vxi) - \vx\cdot\vy
    \ge 0
\label{FYIE-2}
\end{equation}
is interpreted as the Second Law. When $\vx$ and $\vy$ satisfy the relation that validates the equality in Eq. (\ref{FYIE-2}), we say the thermodynamic conjugate variables are in equilibrium.

One of the convincing examples of the thermodynamic duality structures is between relative entropy as the level II large deviation rate function for empirical frequency of independent, identically distributed (i.i.d.) samples, also known as Kullbeck-Leibler divergence \cite{cyq-21}:
\begin{subequations}
\label{KLD}
\begin{equation}
     \Phi(\vnu\|\vp) =
     \sum_{i=1}^K \nu_i\log\left(\frac{\nu_i}{p_i}\right),
\end{equation}
and its LFT
\begin{equation}
     \Psi(\vu) = \sup_{\vnu\in\mathcal{M}}\Big\{\,
       \vnu\cdot\vu -\phi(\vnu\|\vp) \Big\}
      = \log\sum_{i=1}^K p_ie^{u_i},
\end{equation}
\end{subequations}
where $\mathcal{M}$ is the $K-1$ dimensional probability simplex in $\mathbb{R}^K$. The Fenchel-Young inequality in (\ref{FYIE-2}) now takes the form
\begin{eqnarray}
   \eta(\vnu,\vu) &=& \sum_{i=1}^K \nu_i\log\left(\frac{\nu_i}{p_i}\right) + \log\sum_{i=1}^K p_ie^{u_i} - \sum_{i=1}^K \nu_iu_i
\nonumber\\
  &=& \sum_{i=1}^K \nu_i \log\nu_i
  - \sum_{i=1}^K \nu_i\log\left( \frac{ p_ie^{u_i} }{\sum_{j=1}^K p_je^{u_j} }  \right) \ge 0.
  \label{2ndlaw}
\end{eqnarray}
The equilibrium relation between $\vnu$ and $\vu$ is  \cite{qian_jctc}
\begin{equation}
        \nu_i = \frac{p_i e^{u_i}}{
        \displaystyle \sum_{j=1}^K
         p_j e^{u_j} }.
\end{equation}
It is reduced to Boltzmann's relation if one identifies $-\vu$ as energy and takes $p_i=$ constant, widely known as the postulate of {\em equal a priori probability}.

One notices that $p_ie^{u_i}$ appears in Eq. (\ref{2ndlaw}) as a single entity.  This suggests that probability $\vp$ and energetic $\vu$ are two equivalent mathematical representations of the same empirical reality $\vnu$.

\subsection{Conditional expectation, affine transformation and its dual}
\label{sec:2.C}

We again consider $(\Omega,\mathcal{F},\mathbb{P})$ with $\|\Omega\|=K$, $\mathcal{F}=2^{\Omega}$, and $\mathbb{P}=\vp$.  When a random variable $\vX$ is considered as a vector in $\mathbb{R}^K$, conditional expectation w.r.t. $\mathcal{G}$ defines a linear transformation $\mT_{\mathcal{G}}$,
$\mT_{\mathcal{G}}\vX:=\mathbb{E}[\vX|\mathcal{G}]$.  Then $\mathbb{E}\big[\mathbb{E}[\vX|\mathcal{G}]\big]=\mathbb{E}[\vX]$ for all $\vX$ implies the following key properties of $\mT_{\mathcal{G}}$:

(i) $\mT_{\mathcal{G}}=(T_{ij})_{K\times K}$ has non-negative $T_{ij}$ and
\[
      \sum_{j=1}^K T_{ij} = 1,
\]
for all $i=1,\cdots,K$.  $\mT_{\mathcal{G}}$ therefore is an affine transformation, a Markov matrix.

(ii) $\mT_{\mathcal{G}}$ is non-invertible with rank $G$; $\mT_{\mathcal{G}}^2=\mT_{\mathcal{G}}$ is a projection operator.

(iii) $\mT_{\mathcal{G}}^T\, \vp =\vp$ is invariant to the affine transformation.  Equivalently, for any additive set function $\mu$: $\mathcal{F}\to\mathbb{R}$, $\mu\big(\mT_{\mathcal{G}}(A)\big)=\mu(A)$ when $A\in\mathcal{G}\subset\mathcal{F}$.

The information inequality in Sec. \ref{sec:2.A} now takes the form:
\begin{eqnarray}
    \sum_{i=1}^K p_i\phi\Big((\mT_{\mathcal{G}}\vX)_i \Big)  &\le& \sum_{i=1}^K p_i \sum_{j=1}^K
       T_{ij}\phi(x_j)
\\
    &=& \sum_{j=1}^K \left(\sum_{i=1}^K p_i
       T_{ij} \right) \phi(x_j) \ = \
       \sum_{j=1}^K p_j \phi(x_j).
\nonumber
\end{eqnarray}

\section{Markov Dynamics}
\label{sec:markov}

This section establishes the fact that a host of Boltzmann's $H$-theorem like inequalities for general Markov dynamics, which has an underlying $\sigma$-algebra filtration, can also be obtained based on an arbitrary convex function.  Only the Markov trajectory additivity of stochastic entropy production \cite{jqq-book} is critically dependent upon logarithm-based information \cite{shore-johnson}.  Through this analysis, it becomes clear what are the robust features of information inequalities, and what is the specific consequence of the Shannon information entropy.

\subsection{Localized information decreases with time}
\label{sec:3.A}

Consider a finite length Markov chain $X_0,X_1,\cdots,X_N$. We again assume a finite state space $\mathfrak{S}=\{1,\cdots,K\}$.  In the modern theory of probability,
\[
\Omega= \underbrace{\mathfrak{S}\otimes\mathfrak{S}\otimes\cdots\otimes\mathfrak{S} }_{N+1},
\]
$\mathcal{F}=2^{\Omega}$ is the largest $\sigma$-algebra possible, and a Markov measure $\mathbb{P}=\xi_{i_0}p_{i_0i_1}p_{i_1i_2}\cdots p_{i_{N-1}i_N}$ for $\omega=(i_0,i_1,\cdots,i_N)\in\Omega$.  Let us fix a time $n$, $0<n<N$, and consider a random variable $Y_{n} =g(X_{n})$ which is solely determined by the state $X_{n}$. Then one has an important result on conditional expectation of $Y_n$:
\begin{equation}
\mathbb{E}\big[Y_n|X_k,\cdots,X_{\ell}\big]
  = \left\{\begin{array}{ccc}
    \mathbb{E}[Y_n|X_k] &&  n<k<\ell \\
    Y_n &&  k\le n\le\ell \\
    \mathbb{E}[Y_n|X_{\ell} ] && k<\ell<n
       \end{array}\right.
\end{equation}
Suppose $\phi$: $\mathbb{R}\to\mathbb{R}$ is a convex function, then one has
\begin{eqnarray}
     && \mathbb{E}\Big[\phi\Big(\mathbb{E} \big[Y_0\big| X_{\ell} \big] \Big)\Big] = \mathbb{E}\left[\, \mathbb{E}\Big[\phi\Big(\mathbb{E}\big[Y_0\big| X_{\ell}] \Big) \,\Big|\, X_k,\cdots,X_{\ell-1} \Big] \,\right]
\nonumber\\
    &\le& \mathbb{E}\Big[\phi\Big(\mathbb{E} \big[Y_0\big| X_k,\cdots,X_{\ell-1},X_{\ell}\big] \Big) \Big] =
    \mathbb{E}\Big[\phi\Big(\mathbb{E}\big[Y_0\big| X_k \big] \Big) \Big],
\label{y0}
\end{eqnarray}
for $0\le k\le\ell\le N$, and similarly:
\begin{equation}
    \mathbb{E}\Big[\phi\Big( \mathbb{E}
    \big[Y_N\big| X_{\ell} \big]\Big)\Big] \ge
     \mathbb{E}\Big[\phi\Big( \mathbb{E}
    \big[Y_N\big|X_k \big]\Big)\Big].
\label{yN}
\end{equation}
If one identifies
\begin{equation}
    H_{Y_n}(X_k):=\mathbb{E}\Big[ \phi\Big( \mathbb{E}
    \big[Y_n\big|X_k \big]\Big) \Big],
\end{equation}
as the ``information'' on $X_n$ at time $k$, then Eqs. (\ref{y0}) and (\ref{yN}) precisely show that, numerical value aside, the information decreases with increasing $|n-k|$. But there is no ``direction of time''.

Let there be a second probability measure $\mathbb{Q}$ on $(\Omega,\mathcal{F})$. The Radon-Nikodym derivative has the important property:
\begin{eqnarray}
    \mathbb{E}^{\mathbb{P}}\left[\left. \frac{\rd\mathbb{Q}}{\rd\mathbb{P}}\,\right| X_k\right] &=& \frac{\displaystyle \sum_{i_0,\cdots,i_{k-1},i_{k+1},\cdots i_N}
     \mathbb{P}\{i_0,\cdots,i_{k-1},X_k,i_{k+1},\cdots i_N\}
    \frac{\mathbb{Q}\{i_0,\cdots,i_{k-1},X_k,i_{k+1},\cdots i_N\} }{\mathbb{P}\{i_0,\cdots,i_{k-1},X_k,i_{k+1},\cdots i_N\} } }{\displaystyle
    \sum_{i_0,\cdots,i_{k-1},i_{k+1},\cdots i_N}
     \mathbb{P}\{i_0,\cdots,i_{k-1},X_k,i_{k+1},\cdots i_N\}
      }
\nonumber\\
    &=&  \frac{\displaystyle \sum_{i_0,\cdots,i_{k-1},i_{k+1},\cdots i_N}
     \mathbb{Q}\{i_0,\cdots,i_{k-1},X_k,i_{k+1},\cdots i_N\} }{\displaystyle
    \sum_{i_0,\cdots,i_{k-1},i_{k+1},\cdots i_N}
     \mathbb{P}\{i_0,\cdots,i_{k-1},X_k,i_{k+1},\cdots i_N\}
      } = \frac{\mathbb{Q}\{X_k\} }{ \mathbb{P}\{X_k\}}.
\end{eqnarray}
If furthermore $\mathbb{Q}$ only differs from $\mathbb{P}$ by the initial distributions for $X_0$:
$\mathbb{Q}\{X_0=i\}=\zeta_i$, $\mathbb{Q}\{X_{n+1}=j|X_n=i\}=p_{ij}$, then
\begin{equation}
\label{simpleQ}
   \frac{\rd\mathbb{Q}}{\rd\mathbb{P}}
   = \frac{\mathbb{Q}\{X_0\} }{\mathbb{P}\{X_0\}}.
\end{equation}
For convex function $\phi(x)=-\log x$ and applying Eq. (\ref{y0}):
\begin{equation}
    -\mathbb{E}^{\mathbb{P}}\left[\,
    \log\left( \mathbb{E}^{\mathbb{P}}
    \left[\left. \frac{\rd\mathbb{Q}}{\rd\mathbb{P}} \,\right| X_{\ell} \right]\right)\right]
    \le -\mathbb{E}^{\mathbb{P}}\left[
    \log\left( \mathbb{E}^{\mathbb{P}}
    \left[\left. \frac{\rd\mathbb{Q}}{\rd\mathbb{P}} \,\right| X_k \right]\right)\right],
\end{equation}
where $\ell\ge k$, that is,
\begin{equation}
    \mathbb{E}^{\mathbb{P}}\left[ \,
    \log\left( \frac{ \mathbb{P}\{ X_{\ell} \} }{\mathbb{Q}\{ X_{\ell} \} } \right) \right]
    \le \mathbb{E}^{\mathbb{P}}\left[ \,
    \log\left( \frac{ \mathbb{P}\{ X_{k} \} }{\mathbb{Q}\{ X_k \} } \right)
     \right].
\label{fed}
\end{equation}
Eq. (\ref{fed}) is the famous $H$-theorem in Markov dynamics \cite{Ge-Qian-2010,voigt_cmp}, a well-known inequality in information theory \cite{cover-book}.
Its origin resides in Eq. (\ref{simpleQ}).

\subsection{Dynamical information balance equation}
\label{sec:3.B}

The probability measure $\mathbb{Q}$ in Eq. (\ref{simpleQ}) is the simplest alternative to $\mathbb{P}$: They differ only at the initial $X_0$.  The decay of information entropy in (\ref{fed}), therefore, is a natural reflection of the $\sigma$-algebra filtration.  For a general $\mathbb{Q}$,
or a {\em non-local} random variable $Y$ that is a multivariate function of all $X_0,\cdots,X_N$, monotonic change of
\begin{equation}
    \mathbb{E}^{\mathbb{P}}\big[\phi\big(
    \mathbb{E}^{\mathbb{P}}[Y|X_k]\big)]
\end{equation}
with $k$ cannot be established in general. However, its change,
\begin{subequations}
\label{ibe}
\begin{equation}
     \dot{H}_Y(k) :=  \mathbb{E}^{\mathbb{P}}\big[\phi\big(\mathbb{E}^{\mathbb{P}}[Y|X_{k+1}]\big)]- \mathbb{E}^{\mathbb{P}}\big[\phi\big(\mathbb{E}^{\mathbb{P}}[
    Y|X_k]\big)]
\end{equation}
can be decomposed into two non-negative components, the information gain on $Y$, $G_Y(k)$, and the information loss on $Y$, $L_Y(k)$:
\begin{eqnarray}
    &\displaystyle
    \dot{H}_Y = G_{Y}-L_{Y},
\\
    &\displaystyle
    G_Y(k) = \mathbb{E}^{\mathbb{P}}\Big[ \phi\Big( \mathbb{E}^{\mathbb{P}}
    \big[ Y\big|X_k,X_{k+1} \big]\Big)-\phi\Big( \mathbb{E}^{\mathbb{P}}
    \big[Y|X_k \big]\Big) \Big]
    \ge 0,
\\
    &\displaystyle
    L_Y(k) = \mathbb{E}^{\mathbb{P}}\Big[ \phi\Big( \mathbb{E}^{\mathbb{P}}
    \big[Y\big|X_k,X_{k+1}\big]\Big)-\phi\Big( \mathbb{E}^{\mathbb{P}}
    \big[\tfrac{\rd\mathbb{Q}}{\rd\mathbb{P}} \big|X_{k+1} \big]\Big) \Big] \ge 0.
\end{eqnarray}
\end{subequations}
Numerical value which depends on the choices of $\phi$ aside, we shall call Eq. (\ref{ibe}) {\em dynamic information balance equation} for random variable $Y$.

If we assume $\mathbb{Q}$ is absolutely continuous w.r.t. $\mathbb{P}$ and consider $Y=\frac{\rd\mathbb{Q}}{\rd\mathbb{P}}$, then $\phi$-based ``free energy'',
\begin{eqnarray}
\label{eq29}
    \dot{F}_{\phi}:= \mathbb{E}^{\mathbb{P}}\Big[
     \phi\Big( \mathbb{E}^{\mathbb{P}}
    \big[\tfrac{\rd\mathbb{Q}}{\rd\mathbb{P}} \big|X_{k+1} \big]\Big) -
    \phi\Big( \mathbb{E}^{\mathbb{P}}
    \big[\tfrac{\rd\mathbb{Q}}{\rd\mathbb{P}} \big|X_k \big]\Big) \Big] =
    Q_{hk}-e_p,
\end{eqnarray}
where house-keeping heat $Q_{hk}$ and entropy production $e_p$ \cite{Ge-Qian-2010}:
\begin{subequations}
\begin{eqnarray}
    &\displaystyle
    Q_{hk} = \mathbb{E}^{\mathbb{P}}\Big[ \phi\Big( \mathbb{E}^{\mathbb{P}}
    \big[ \tfrac{\rd\mathbb{Q}}{\rd\mathbb{P}}\big|X_k,X_{k+1} \big]\Big)-\phi\Big( \mathbb{E}^{\mathbb{P}}
    \big[\tfrac{\rd\mathbb{Q}}{\rd\mathbb{P}}|X_k \big]\Big) \Big]
    \ge 0,
\\
    &\displaystyle
     e_p = \mathbb{E}^{\mathbb{P}}\Big[ \phi\Big( \mathbb{E}^{\mathbb{P}}
    \big[\tfrac{\rd\mathbb{Q}}{\rd\mathbb{P}}\big|X_k,X_{k+1}\big]\Big) -\phi\Big( \mathbb{E}^{\mathbb{P}}
    \big[\tfrac{\rd\mathbb{Q}}{\rd\mathbb{P}} \big|X_{k+1} \big]\Big) \Big] \ge 0.
\end{eqnarray}
\end{subequations}
There is a very general stochastic $\phi$-based free energy balance.

If $\mathbb{Q}$ is another Markovian measure with
\[
  \mathbb{Q}\big\{X_0=i_0,\cdots,X_N=i_N\big\}
    =\zeta_{i_0} q_{i_0i_1} q_{i_1i_2} \cdots
      q_{i_{N-1}i_N},
\]
and $q_{ij}=\pi_jp_{ji}/\pi_i$ where $\pi_i$ is the stationary distribution of $p_{ij}$, then
\begin{equation}
\mathbb{E}^{\mathbb{P}}
    \Big[\tfrac{\rd\mathbb{Q}}{\rd\mathbb{P}}\Big|X_k,X_{k+1}\Big]
    = \frac{\mathbb{Q}\{X_k,X_{k+1}\}}{\mathbb{P}\{X_k,X_{k+1}\}} =
     \frac{q_{X_kX_{k+1}}}{p_{X_kX_{k+1} }} \, \mathbb{E}^{\mathbb{P}}\Big[
     \tfrac{\rd\mathbb{Q}}{\rd\mathbb{P}} \Big| X_k \Big],
\end{equation}
and
\begin{equation}
        Q_{hk} = \mathbb{E}^{\mathbb{P}}\Big[ \phi\Big( \tfrac{\pi_{X_{k+1}}p_{X_{k+1}X_k}}{\pi_{X_k}p_{X_kX_{k+1} }} \, \mathbb{E}^{\mathbb{P}}\big[
     \tfrac{\rd\mathbb{Q}}{\rd\mathbb{P}} \big| X_k \big] \Big)-\phi\Big( \mathbb{E}^{\mathbb{P}}
    \big[\tfrac{\rd\mathbb{Q}}{\rd\mathbb{P}}|X_k \big]\Big) \Big].
\end{equation}
$Q_{hk}$ is zero if the Markov chain is reversible; its $\dot{F}_{\phi}=-e_p\le 0$.

Eq. (\ref{eq29}) shows that Markov process can be further refined with the notion of reversibility, or detailed balance: $\pi_ip_{ij}=\pi_jp_{ji}$ \cite{jqq-book}.  For reversible Markov chain, there is a special $\mathbb{Q}$ under which $F_{\phi}$ becomes the $H$ in Sec. \ref{sec:3.A}.  For an irreversible Markov chain, its stationary process has equal non-zero $e_p$ and $Q_{hk}$: It is sustained by a driving force and a dissipation \cite{Ge-Qian-2010}. This is one of the insights of the Brussels school of nonequilibrium thermodynamics \cite{prigogine-book,jqq-book}

The very fact that the mathematical relations in Eqs. (\ref{y0}), (\ref{yN}), and (\ref{ibe}) are true for arbitrary convex function $\phi$ illustrates that the information inequality is a topological feature hidden in the $\sigma$-algebra \cite{urbanik}; it is merely being brought out by a convex function \cite{rockafellar-book}.

\subsection{Global trajectory-based entropy production}

Since a probability measure is defined on the $(\Omega,\mathcal{F})$, one can introduce a $\tilde{\mathbb{Q}}$ more globally as
\begin{eqnarray}
   \tilde{\mathbb{Q}}\big\{X_0=i_0,\cdots,X_N=i_N\big\}
    &=& \mathbb{P}\big\{X_0=i_N,X_1=i_{N-1}\cdots,X_N=i_0\big\}
\nonumber\\
    &=& \xi_{i_N}p_{i_Ni_{N-1}}p_{i_{N-1}i_{N-2}}\cdots p_{i_2i_1}p_{i_1i_0}.
\end{eqnarray}
$\tilde{\mathbb{Q}}$ is again a Markov measure: $\tilde{\mathbb{Q}}\{X_k=i_k|X_{k-1},X_{k-2},\cdots,X_0\}= \tilde{\mathbb{Q}}\{X_k=i_k|X_{k-1} \}$.  However it is no longer time homogeneous in general; the only exception is when $\xi_i=\pi_i$.

For $\omega=(i_0,\cdots,i_N)$,
\begin{eqnarray}
    \frac{\rd\tilde{\mathbb{Q}}}{\rd\mathbb{P}}(\omega)
    &=& \frac{ \tilde{\mathbb{Q}}\big\{X_0=i_0,\cdots,X_N=i_N\big\} }{ \mathbb{P}\big\{X_0=i_0,\cdots,X_N=i_N\big\}  } =
    \frac{\xi_{i_N}}{\xi_{i_0}}\prod_{\ell=0}^{N-1}
    \frac{ p_{i_{\ell+1}i_{\ell} } }{ p_{i_{\ell}i_{\ell+1} } },
\label{eq31}
\end{eqnarray}
its logarithm has an additivity along the trajectory:
\begin{equation}
     \log\left(\frac{\rd\tilde{\mathbb{Q}}}{\rd\mathbb{P}}(\omega)\right)
     = \sum_{\ell=0}^{N-1} \log\left(\frac{ \xi_{i_{\ell+1}}p_{i_{\ell+1}i_{\ell} } }{ \xi_{i_{\ell}}p_{i_{\ell}i_{\ell+1} } } \right) = \sum_{\ell=0}^{N-1}
    \log\left(\mathbb{E}^{\mathbb{P}}\left[  \left. \frac{\rd\tilde{\mathbb{Q}}}{\rd\mathbb{P}}\right| X_{\ell},X_{\ell+1} \right]  \right).
\end{equation}
Using the logarithmic convex function and identifying Eq. (\ref{eq31}) as the $N$-step stochastic entropy production of a random $\omega$, then it is the sum of the logarithm-based entropy production of each individual Markov step.  This additivity is a defining feature of logarithm-based stochastic information entropy.

\section{Coarse-Grained Thermodynamic Effective Theory}
\label{sec:cg}

With the Fenchel-Young inequality based on the pair of dual convex functions $\phi(\vx)$ and $\psi(\vy)$ on a thermo-doubled space, we now identify
\begin{equation}
\label{fyi-2}
    \eta(\vx,\vy) :=
    \phi(\vx)+\psi(\vy)-\vx\cdot\vy
    \ge 0, \ \ (\vx,\vy)\in\mathbb{R}^K\otimes\mathbb{R}^K
\end{equation}
as {\em entropy production}, {\em \`{a} la} the Brussels school of nonequilibrium thermodynamics \cite{prigogine-book}.  The $K$-dimensional equilibrium manifold embodied by $\eta(\vx,\vy)=0$ then is captured by a bijective relation between $\vx$ and its conjugate $\vy^{\text{eq}}(\vx)=\nabla\phi(\vx)$, or equivalently $\vx^{\text{eq}}(\vy)=\nabla\psi(\vy)$.

Under an invertible linear transformation $\mT$: LFT has
\begin{equation}
   \psi(\mT\vy)= \sup_{\vx\in\mathbb{R}^K}
   \Big\{ \vx\cdot\mT\vy - \phi(\vx) \Big\}
   = \sup_{\vx'\in\mathbb{R}^K}
   \Big\{ \vx'\cdot\vy - \phi\big(\mT^{*-1}\vx'\big) \Big\},
\end{equation}
which corresponds to $\phi(\mT^{*-1}\vx)$. With $\vy'=\mT\vy$ and $\vx'=\mT^{*-1}\vx$, $(\vy')^{\text{eq}}(\vx)=\mT\nabla_{\vx}\phi(\mT^{*-1}\vx)$. The spaces of $\vx$ and $\vy$ are not only dual, but also reciprocal. Eq. (\ref{fyi-2}) becomes
$\phi(\mT^*\vx)+\psi(\mT^{-1}\vy)-\vx\cdot\vy\ge 0$:
While $\phi(\vx)\to\phi(\mT^*\vx)$ and
corresponding $\psi(\vy)\to\psi(\mT^{-1}\vy)$, $(\mT^*\vx)\cdot(\mT^{-1}\vy)=\vx\cdot\vy$ is unchanged.  This fits with identifying the three terms as ``entropy $+$ free energy $-$ internal energy''.

Coarse-graining is represented by a non-invertible transformation $\mT$, and through LFT one sees that ``projection'' and/or ``constraint'' are two different perspectives on a same transformation.  Let $\vy'=\mT\vy$ be a description of a system with lower resolution: $\mT$ maps many different $\vy$'s to a same $\vy'$ in a linear sub-space, the range $\mathcal{R}(\mT)\subset\mathbb{R}^K$.  One concrete example of $\mT$ is the conditional expectation discussed in Sec. \ref{sec:2.C}.

We emphasize the distinction between $\tilde{\psi}(\vy):=\psi(\mT\vy)$ which is defined on the entire $\mathbb{R}^K$ and $\psi(\vy')$ with $\vy'\in\mathcal{R}(\mT)$: The former is not a convex function since it has equal value for all $\vy$ with $\mT\vy=\vy'$.  Under the $\mT$, a new Frenchel-Young equilibrium equality
\begin{equation}
\label{eq36}
    \tilde{\eta}(\vx,\vy) :=
    \tilde{\phi}(\vx)+\tilde{\psi}(\vy)-\vx\cdot\vy=0,
\end{equation}
appears, with a pair of new functions $\tilde{\phi}(\vx)$ and $\tilde{\psi}(\vy)$, $(\vx,\vy)\in\mathbb{R}^K\otimes\mathbb{R}^K$, in which
\begin{eqnarray}
    \tilde{\phi}(\vx) &=& \sup_{\vy\in\mathbb{R}^K}\Big\{\vx\cdot\vy
 - \tilde{\psi}(\vy) \Big\}
\nonumber\\
    &=& \sup_{\vy\in\mathbb{R}^K}\Big\{\vx\cdot\vy
 -\psi(\mT\vy) \Big\}
\nonumber\\
    &=& \left\{\begin{array}{ccc}
      \displaystyle
      \inf_{\vx'\in\mathbb{R}^K} \big\{\phi(\vx') \,\big|\,
     \mT^*\vx' =\vx\big\} && \vx\in \mathcal{R}(\mT^*)  \\
      \infty   &&    \text{otherwise}
      \end{array}\right.
\label{eq37}
\end{eqnarray}
The restriction on the support of $\tilde{\phi}(\vx)$ is a consequence of the non-convexity of $\tilde{\psi}(\vy)$ for $\vy\in\mathbb{R}^K$.  To see the $\infty$ in (\ref{eq37}), we denote $\vy=\va+\vb$ with $\va=\mathcal{R}(\mT^*)$, $\vb=\mathcal{N}(\mT)$, and $\va\cdot\vb=0$.  Then $\tilde{\psi}(\vy)=\psi\big(\mT(\va+\vb)\big)=\psi(\mT\va)$ is independent of $\vb$. When $\vx\in\mathcal{R}(\mT^*)$, $\vx\cdot\vy=\vx\cdot(\va+\vb)=\vx\cdot\va$.  However, when $\vx\notin\mathcal{R}(\mT^*)$, $\vb\neq {\bf 0}$, $\vx\cdot\vb$ can be arbitrarily small when $\|\vb\|\to\infty$.  For $\vx$ restricted on $\mathcal{R}(\mT^*)$:
\begin{eqnarray}
 \tilde{\phi}(\vx) &=& \sup_{\vy\in\mathbb{R}^K}\Big\{\vx\cdot\vy
 -\psi(\mT\vy) \Big\}
\nonumber\\
    &=& \sup_{\vy\in\mathbb{R}^K}\Big\{\vx\cdot\vy
 - \sup_{\vx'\in\mathbb{R}^K} \big\{ \vx'\cdot\mT\vy -\phi(\vx') \big\} \Big\}
\nonumber\\
    &=& \sup_{\vy\in\mathbb{R}^K}\Big\{\vx\cdot\vy
  + \inf_{\vx'\in\mathbb{R}^K} \big\{ \phi(\vx')-\vx'\cdot\mT\vy \big\} \Big\}
\nonumber\\
    &=& \inf_{\vx'\in\mathbb{R}^K} \Big\{\phi(\vx')
  + \sup_{\vy\in\mathbb{R}^K} \big\{\big(\vx-\mT^*\vx'\big)\cdot\vy \big\} \Big\}
\nonumber\\
    &=& \inf_{\vx'\in\mathbb{R}^K} \Big\{\phi(\vx') \,\Big|\,
     \mT^*\vx' =\vx \Big\}.
\label{eq41}
\end{eqnarray}
Under a non-invertible transformation $\mT$, an ``effective theory'' appears with the corresponding free energy $\tilde{\psi}(\vy)=\psi(\mT\vy)$ and entropy function $\tilde{\phi}(\vx)$ obtained from constrained minimization of $\phi(\vx')$, $\mT^*\vx'=\vx$.
This mathematical result reflects a deep connection between statistical ensemble change in Gibbs' theory and macroscopic thermodynamics; see Appendix \ref{appendix}.

Formally Eq. (\ref{eq36}) is still defined on the entire $(\vx,\vy)\in\mathbb{R}^K\otimes\mathbb{R}^K$ on which $\tilde{\phi}(\vx)$ and $\tilde{\psi}(\vy)$ are not convex functions. In actuality the support of
finite $\tilde{\phi}(\vx)$ is $\mathcal{R}(\mT^*)$, $\vx=\mT^*\vx'$.  Similarly, restricting $\vy\in\mathcal{R}(\mT^*)$, $\tilde{\psi}(\vy)=\psi(\mT\vy)$ is a convex function.  Eq. (\ref{eq36}) with $(\vx,\vy)\in \mathcal{R}(\mT^*)\otimes\mathcal{R}(\mT^*)$ regains equilibrium duality symmetry for the effective theory.

\section{Discussion}
\label{sec:5}

Coarse-grained representation and/or incomplete measurements of a system are processes that involve information reduction. In the modern theory of probability, they are modelled through {\em conditioning} on sub-$\sigma$-algebra, which is a more rigorous formulation of the idea of ``partial averaging''.  The value of a convex function always decreases under pre-average. Convex functions therefore can be employed to express information reduction.  The Shannon entropy and its variants are simply a special class of convex (or concave) functions.

Thermodynamic theory balances a convex function $\phi(\vx)$, $\vx\in\mathbb{R}^K$, by a conjugate function $\psi(\vy)$ through LFT.  The $\phi(\vx)$ and $\psi(\vy)$ together in the thermo-doubled space of $(\vx,\vy)\in\mathbb{R}^K\otimes\mathbb{R}^K$ defines an equilibrium relationship between $\vx$ and $\vy$ through $\phi(\vx)+\psi(\vy)-\vx\cdot\vy=0$.  Since $\phi(\vx)$ can be considered as a thermodynamic potential function, the equilibrium $\vy^{\text{eq}}=\nabla\phi(\vx)$, is naturally called {\em thermodynamic force}.  For all other nonequilibrium state $(\vx,\vy)$, $\eta(\vx,\vy):=\phi(\vx)+\psi(\vy)-\vx\cdot\vy > 0$ captures an irreversible tendency.
The idea of space doubling has its inspiration in the earlier work of Schwinger, Keldysh, and Martin-Siggia-Rose \cite{schwinger-61,keldysh-65,MSR-73}; the present work makes it fundamental that {\em equilibrium} is between $\vx$ and its conjugate $\vy$ via Fenchel-Young equality in the thermo-doubled space.

So far the discussion is for an arbitrary convex function.  When a particular set of observables (random variables) with measurement data {\em ad infinitum}, a convex function that is intrinsic to the probabilistic system and its measurement emerges. This is known as {\em large deviations theory} in mathematics. In terms of this particular convex function, one quantifies the amount of information in a measurement w.r.t. a statistical model.  One also quantifies the goodness of a model w.r.t. the empirical observations.  Maximum likelihood principle appears in the latter context.

When time goes on, there are ``more information''.  Therefore, time, information, and entropy are forever bound together \cite{thermo_info}.  The present work reveals a deeper common mathematical origin of these concepts.  Interestingly, as shown in Eqs. (\ref{y0}) and (\ref{yN}), when a Markov process is conditioned on $X_N$ at time $N$, {\em i.e.} the time $N$ has already passed, mathematics shows that the arrow of time is lost: Only the time interval matters.  Indeed, the large deviations theory for Markov dynamics and analytical mechanics share the same mathematical structure \cite{ge_qian_ijmpb}.

More advanced mathematics is needed when extending the current logic for $\mathbb{R}^K$, the space of all random variables on a finite $\Omega$ with $\|\Omega\|=K$, to continuous $\Omega$ with infinite-dimensional function space $\mathcal{V}(\Omega)$ whose algebraic dual, the space of measures, always has a larger cardinal number than $\mathcal{V}$'s.  An appropriate Banach space of bounded continuous functions, whose dual are measures with density functions, is needed for establishing the duality in thermodynamic equilibrium.  Self-adjoint symmetry then is further formulated in a Hilbert space \cite{jqq-book,qqt_jsp}.

{\bf Acknowledgement. }
The authors thank Professor Quanhui Liu (Hunan Univ.) for continuous encouragement and the members of Online International Club Nanothermodynamica for discussions.  H. Q. was partially supported by the Olga Jung Wan Endowed Professorship.

%

\bibliography{reference}

\begin{thebibliography}{38}%
\makeatletter
\providecommand \@ifxundefined [1]{%
 \@ifx{#1\undefined}
}%
\providecommand \@ifnum [1]{%
 \ifnum #1\expandafter \@firstoftwo
 \else \expandafter \@secondoftwo
 \fi
}%
\providecommand \@ifx [1]{%
 \ifx #1\expandafter \@firstoftwo
 \else \expandafter \@secondoftwo
 \fi
}%
\providecommand \natexlab [1]{#1}%
\providecommand \enquote  [1]{``#1''}%
\providecommand \bibnamefont  [1]{#1}%
\providecommand \bibfnamefont [1]{#1}%
\providecommand \citenamefont [1]{#1}%
\providecommand \href@noop [0]{\@secondoftwo}%
\providecommand \href [0]{\begingroup \@sanitize@url \@href}%
\providecommand \@href[1]{\@@startlink{#1}\@@href}%
\providecommand \@@href[1]{\endgroup#1\@@endlink}%
\providecommand \@sanitize@url [0]{\catcode `\\12\catcode `\$12\catcode `\&12\catcode `\#12\catcode `\^12\catcode `\_12\catcode `\%12\relax}%
\providecommand \@@startlink[1]{}%
\providecommand \@@endlink[0]{}%
\providecommand \url  [0]{\begingroup\@sanitize@url \@url }%
\providecommand \@url [1]{\endgroup\@href {#1}{\urlprefix }}%
\providecommand \urlprefix  [0]{URL }%
\providecommand \Eprint [0]{\href }%
\providecommand \doibase [0]{http://dx.doi.org/}%
\providecommand \selectlanguage [0]{\@gobble}%
\providecommand \bibinfo  [0]{\@secondoftwo}%
\providecommand \bibfield  [0]{\@secondoftwo}%
\providecommand \translation [1]{[#1]}%
\providecommand \BibitemOpen [0]{}%
\providecommand \bibitemStop [0]{}%
\providecommand \bibitemNoStop [0]{.\EOS\space}%
\providecommand \EOS [0]{\spacefactor3000\relax}%
\providecommand \BibitemShut  [1]{\csname bibitem#1\endcsname}%
\let\auto@bib@innerbib\@empty
\bibitem [{\citenamefont {Cover}\ and\ \citenamefont {Thomas}(2006)}]{cover-book}%
  \BibitemOpen
  \bibfield  {author} {\bibinfo {author} {\bibfnamefont {T.~M.}\ \bibnamefont {Cover}}\ and\ \bibinfo {author} {\bibfnamefont {J.~A.}\ \bibnamefont {Thomas}},\ }\href@noop {} {\emph {\bibinfo {title} {Elements of Information Theory}}},\ \bibinfo {edition} {2nd}\ ed.\ (\bibinfo  {publisher} {Wiley-Interscience},\ \bibinfo {address} {New York},\ \bibinfo {year} {2006})\BibitemShut {NoStop}%
\bibitem [{\citenamefont {Hobson}(1969)}]{hobson_1969}%
  \BibitemOpen
  \bibfield  {author} {\bibinfo {author} {\bibfnamefont {A.}~\bibnamefont {Hobson}},\ }\bibfield  {title} {\enquote {\bibinfo {title} {A new theorem of information theory},}\ }\href {\doibase 10.1007/BF01106578} {\bibfield  {journal} {\bibinfo  {journal} {Journal of Statistical Physics}\ }\textbf {\bibinfo {volume} {1}},\ \bibinfo {pages} {383--391} (\bibinfo {year} {1969})}\BibitemShut {NoStop}%
\bibitem [{\citenamefont {(ed.)}(2003)}]{karmeshu-book}%
  \BibitemOpen
  \bibfield  {author} {\bibinfo {author} {\bibfnamefont {Karmeshu}\ \bibnamefont {(ed.)}},\ }\href {\doibase 10.1007/978-3-540-36212-8} {\emph {\bibinfo {title} {Entropy Measures, Maximum Entropy Principle and Emerging Applications}}}\ (\bibinfo  {publisher} {Springer},\ \bibinfo {address} {Berlin},\ \bibinfo {year} {2003})\BibitemShut {NoStop}%
\bibitem [{\citenamefont {Peng}\ \emph {et~al.}(2020)\citenamefont {Peng}, \citenamefont {Qian},\ and\ \citenamefont {Hong}}]{hongliu_tsallis}%
  \BibitemOpen
  \bibfield  {author} {\bibinfo {author} {\bibfnamefont {L.}~\bibnamefont {Peng}}, \bibinfo {author} {\bibfnamefont {H.}~\bibnamefont {Qian}}, \ and\ \bibinfo {author} {\bibfnamefont {L.}~\bibnamefont {Hong}},\ }\bibfield  {title} {\enquote {\bibinfo {title} {Thermodynamics of {M}arkov processes with nonextensive entropy and free energy},}\ }\href {\doibase 10.1103/PhysRevE.101.022114} {\bibfield  {journal} {\bibinfo  {journal} {Phys. Rev. E}\ }\textbf {\bibinfo {volume} {101}},\ \bibinfo {pages} {022114} (\bibinfo {year} {2020})}\BibitemShut {NoStop}%
\bibitem [{\citenamefont {Kolmogoroff}(1933)}]{KolmogorovBook}%
  \BibitemOpen
  \bibfield  {author} {\bibinfo {author} {\bibfnamefont {A.~N.}\ \bibnamefont {Kolmogoroff}},\ }\href {\doibase 10.1007/978-3-642-49888-6} {\emph {\bibinfo {title} {Grundbegriffe der {W}ahrscheinlichkeitsrechnung}}}\ (\bibinfo  {publisher} {Springer},\ \bibinfo {address} {New York},\ \bibinfo {year} {1933})\BibitemShut {NoStop}%
\bibitem [{\citenamefont {Durrett}(2019)}]{DurrettBook}%
  \BibitemOpen
  \bibfield  {author} {\bibinfo {author} {\bibfnamefont {R.}~\bibnamefont {Durrett}},\ }\href {\doibase 10.1017/9781108591034} {\emph {\bibinfo {title} {Probability Theory and Examples}}},\ \bibinfo {edition} {5th}\ ed.\ (\bibinfo  {publisher} {Cambridge Univ. Press},\ \bibinfo {address} {London},\ \bibinfo {year} {2019})\BibitemShut {NoStop}%
\bibitem [{\citenamefont {Urbanik}(1973)}]{urbanik}%
  \BibitemOpen
  \bibfield  {author} {\bibinfo {author} {\bibfnamefont {K.}~\bibnamefont {Urbanik}},\ }\bibfield  {title} {\enquote {\bibinfo {title} {On the definition of information},}\ }\href {\doibase 10.1016/0034-4877(73)90004-9} {\bibfield  {journal} {\bibinfo  {journal} {Report on Mathematical Physics}\ }\textbf {\bibinfo {volume} {4}},\ \bibinfo {pages} {289--301} (\bibinfo {year} {1973})}\BibitemShut {NoStop}%
\bibitem [{\citenamefont {Lu}\ and\ \citenamefont {Qian}(2022)}]{lu-qian-22}%
  \BibitemOpen
  \bibfield  {author} {\bibinfo {author} {\bibfnamefont {Z.}~\bibnamefont {Lu}}\ and\ \bibinfo {author} {\bibfnamefont {H.}~\bibnamefont {Qian}},\ }\bibfield  {title} {\enquote {\bibinfo {title} {Emergence and breaking of duality symmetry in thermodynamic behavior: {R}epeated measurements and macroscopic limit},}\ }\href {\doibase 10.1103/PhysRevLett.128.150603} {\bibfield  {journal} {\bibinfo  {journal} {Physical Review Letters}\ }\textbf {\bibinfo {volume} {128}},\ \bibinfo {pages} {150603} (\bibinfo {year} {2022})}\BibitemShut {NoStop}%
\bibitem [{\citenamefont {Qian}(2022)}]{qian_jctc}%
  \BibitemOpen
  \bibfield  {author} {\bibinfo {author} {\bibfnamefont {H.}~\bibnamefont {Qian}},\ }\bibfield  {title} {\enquote {\bibinfo {title} {Statistical chemical thermodynamics and energetic behavior of counting: {G}ibbs’ theory revisited},}\ }\href {\doibase 10.1021/acs.jctc.2c00783} {\bibfield  {journal} {\bibinfo  {journal} {J. Chem. Theory Comput.}\ }\textbf {\bibinfo {volume} {18}},\ \bibinfo {pages} {6421--6436} (\bibinfo {year} {2022})}\BibitemShut {NoStop}%
\bibitem [{\citenamefont {Prigogine}(1947)}]{prigogine-book}%
  \BibitemOpen
  \bibfield  {author} {\bibinfo {author} {\bibfnamefont {I.}~\bibnamefont {Prigogine}},\ }\href@noop {} {\emph {\bibinfo {title} {Etude Thermodynamique des Ph\'{e}nom\`{e}nes Irr\'{e}versibles}}}\ (\bibinfo  {publisher} {Dunod},\ \bibinfo {address} {Paris},\ \bibinfo {year} {1947})\BibitemShut {NoStop}%
\bibitem [{\citenamefont {Lieb}\ and\ \citenamefont {Yngvason}(2000)}]{lieb}%
  \BibitemOpen
  \bibfield  {author} {\bibinfo {author} {\bibfnamefont {E.~H.}\ \bibnamefont {Lieb}}\ and\ \bibinfo {author} {\bibfnamefont {J.}~\bibnamefont {Yngvason}},\ }\bibfield  {title} {\enquote {\bibinfo {title} {A fresh look at entropy and the second law of thermodynamics},}\ }\href {\doibase 10.1063/1.883034} {\bibfield  {journal} {\bibinfo  {journal} {Physics Today}\ }\textbf {\bibinfo {volume} {53(4)}},\ \bibinfo {pages} {32--37} (\bibinfo {year} {2000})}\BibitemShut {NoStop}%
\bibitem [{\citenamefont {Yang}\ and\ \citenamefont {Qian}(2022)}]{yangqian_22}%
  \BibitemOpen
  \bibfield  {author} {\bibinfo {author} {\bibfnamefont {Y.-J.}\ \bibnamefont {Yang}}\ and\ \bibinfo {author} {\bibfnamefont {H.}~\bibnamefont {Qian}},\ }\bibfield  {title} {\enquote {\bibinfo {title} {Statistical thermodynamics and data infinitum: {C}onjugate variables as forces, and their statistical variations},}\ }\href {\doibase 10.48550/arXiv.2205.09321} {\bibfield  {journal} {\bibinfo  {journal} {arXiv:2205.09321}\ } (\bibinfo {year} {2022}),\ 10.48550/arXiv.2205.09321}\BibitemShut {NoStop}%
\bibitem [{\citenamefont {Dembo}\ and\ \citenamefont {Zeitouni}(1998)}]{dembo-book}%
  \BibitemOpen
  \bibfield  {author} {\bibinfo {author} {\bibfnamefont {A.}~\bibnamefont {Dembo}}\ and\ \bibinfo {author} {\bibfnamefont {O.}~\bibnamefont {Zeitouni}},\ }\href {\doibase 10.1007/978-3-642-03311-7} {\emph {\bibinfo {title} {Large Deviations Techniques and Applications}}},\ \bibinfo {edition} {2nd}\ ed.\ (\bibinfo  {publisher} {Springer},\ \bibinfo {address} {New York},\ \bibinfo {year} {1998})\BibitemShut {NoStop}%
\bibitem [{\citenamefont {Virinchi}\ \emph {et~al.}(2023)\citenamefont {Virinchi}, \citenamefont {Angelini}, \citenamefont {Gao}, \citenamefont {Hong}, \citenamefont {Hu}, \citenamefont {Sun}, \citenamefont {Thompson},\ and\ \citenamefont {Qian}}]{paper-I}%
  \BibitemOpen
  \bibfield  {author} {\bibinfo {author} {\bibfnamefont {V.~M.}\ \bibnamefont {Virinchi}}, \bibinfo {author} {\bibfnamefont {E.}~\bibnamefont {Angelini}}, \bibinfo {author} {\bibfnamefont {Y.}~\bibnamefont {Gao}}, \bibinfo {author} {\bibfnamefont {L.}~\bibnamefont {Hong}}, \bibinfo {author} {\bibfnamefont {J.}~\bibnamefont {Hu}}, \bibinfo {author} {\bibfnamefont {W.}~\bibnamefont {Sun}}, \bibinfo {author} {\bibfnamefont {L.~F.}\ \bibnamefont {Thompson}}, \ and\ \bibinfo {author} {\bibfnamefont {H.}~\bibnamefont {Qian}},\ }\bibfield  {title} {\enquote {\bibinfo {title} {Statistical foundation of holographic information entropy for data {\it ad infinitum}},}\ }\href@noop {} {\bibfield  {journal} {\bibinfo  {journal} {manuscript in preparation}\ } (\bibinfo {year} {2023})}\BibitemShut {NoStop}%
\bibitem [{\citenamefont {Jaynes}(2003)}]{JaynesBook}%
  \BibitemOpen
  \bibfield  {author} {\bibinfo {author} {\bibfnamefont {E.~T.}\ \bibnamefont {Jaynes}},\ }\href@noop {} {\emph {\bibinfo {title} {Probability Theory: {T}he Logic of Science}}}\ (\bibinfo  {publisher} {Cambridge Univ. Press},\ \bibinfo {address} {London},\ \bibinfo {year} {2003})\BibitemShut {NoStop}%
\bibitem [{\citenamefont {Peliti}\ and\ \citenamefont {Pigolotti}(2021)}]{peliti-book}%
  \BibitemOpen
  \bibfield  {author} {\bibinfo {author} {\bibfnamefont {L.}~\bibnamefont {Peliti}}\ and\ \bibinfo {author} {\bibfnamefont {S.}~\bibnamefont {Pigolotti}},\ }\href@noop {} {\emph {\bibinfo {title} {Stochastic Thermodynamics: An Introduction}}}\ (\bibinfo  {publisher} {Princeton Univ. Press},\ \bibinfo {address} {Princeton, NJ},\ \bibinfo {year} {2021})\BibitemShut {NoStop}%
\bibitem [{\citenamefont {Shiraishi}(2023)}]{shiraishi-book}%
  \BibitemOpen
  \bibfield  {author} {\bibinfo {author} {\bibfnamefont {N.}~\bibnamefont {Shiraishi}},\ }\href {\doibase 10.1007/978-981-19-8186-9} {\emph {\bibinfo {title} {An Introduction to Stochastic Thermodynamics: From Basic to Advanced}}}\ (\bibinfo  {publisher} {Springer Nature},\ \bibinfo {address} {Singapore},\ \bibinfo {year} {2023})\BibitemShut {NoStop}%
\bibitem [{\citenamefont {Qian}(2001{\natexlab{a}})}]{qian_pre_eec}%
  \BibitemOpen
  \bibfield  {author} {\bibinfo {author} {\bibfnamefont {H.}~\bibnamefont {Qian}},\ }\bibfield  {title} {\enquote {\bibinfo {title} {Mesoscopic nonequilibrium thermodynamics of single macromolecules and dynamic entropy-energy compensation},}\ }\href {\doibase 10.1103/PhysRevE.65.016102} {\bibfield  {journal} {\bibinfo  {journal} {Phys. Rev. E}\ }\textbf {\bibinfo {volume} {65}},\ \bibinfo {pages} {016102} (\bibinfo {year} {2001}{\natexlab{a}})}\BibitemShut {NoStop}%
\bibitem [{\citenamefont {Qian}\ \emph {et~al.}(1991)\citenamefont {Qian}, \citenamefont {Qian},\ and\ \citenamefont {Gong}}]{qqg-contemp}%
  \BibitemOpen
  \bibfield  {author} {\bibinfo {author} {\bibfnamefont {M.-P.}\ \bibnamefont {Qian}}, \bibinfo {author} {\bibfnamefont {M.}~\bibnamefont {Qian}}, \ and\ \bibinfo {author} {\bibfnamefont {G.-L.}\ \bibnamefont {Gong}},\ }\bibfield  {title} {\enquote {\bibinfo {title} {The reversibility and the entropy production of {M}arkov processes},}\ }\href@noop {} {\bibfield  {journal} {\bibinfo  {journal} {Contemporary Mathematics}\ }\textbf {\bibinfo {volume} {118}},\ \bibinfo {pages} {157--168} (\bibinfo {year} {1991})}\BibitemShut {NoStop}%
\bibitem [{\citenamefont {Kolmogorov}(1968)}]{kolmogorov-information}%
  \BibitemOpen
  \bibfield  {author} {\bibinfo {author} {\bibfnamefont {A.~N.}\ \bibnamefont {Kolmogorov}},\ }\bibfield  {title} {\enquote {\bibinfo {title} {Three approaches to the quantitative definition of information},}\ }\href {\doibase 10.1080/00207166808803030} {\bibfield  {journal} {\bibinfo  {journal} {International Journal of Computer Mathematics}\ }\textbf {\bibinfo {volume} {2}},\ \bibinfo {pages} {157--168} (\bibinfo {year} {1968})}\BibitemShut {NoStop}%
\bibitem [{\citenamefont {Tribus}(1961)}]{tribus-book}%
  \BibitemOpen
  \bibfield  {author} {\bibinfo {author} {\bibfnamefont {M.}~\bibnamefont {Tribus}},\ }\href@noop {} {\emph {\bibinfo {title} {Thermodynamics and Thermostatics: An Introduction to Energy, Information and States of Matter, with Engineering Applications}}}\ (\bibinfo  {publisher} {D. van Nostrand},\ \bibinfo {address} {New York},\ \bibinfo {year} {1961})\BibitemShut {NoStop}%
\bibitem [{\citenamefont {Jarzynski}(2011)}]{jarzynski}%
  \BibitemOpen
  \bibfield  {author} {\bibinfo {author} {\bibfnamefont {C.}~\bibnamefont {Jarzynski}},\ }\bibfield  {title} {\enquote {\bibinfo {title} {Equalities and inequalities: {I}rreversibility and the second law of thermodynamics at the nanoscale},}\ }\href {\doibase 10.1146/annurev-conmatphys-062910-140506} {\bibfield  {journal} {\bibinfo  {journal} {Annual Review of Condensed Matter Physics}\ }\textbf {\bibinfo {volume} {2}},\ \bibinfo {pages} {329--351} (\bibinfo {year} {2011})}\BibitemShut {NoStop}%
\bibitem [{\citenamefont {Qian}(2001{\natexlab{b}})}]{qian_prsa}%
  \BibitemOpen
  \bibfield  {author} {\bibinfo {author} {\bibfnamefont {H.}~\bibnamefont {Qian}},\ }\bibfield  {title} {\enquote {\bibinfo {title} {Mathematical formalism for isothermal linear irreversibility},}\ }\href {\doibase 10.1098/rspa.2001.0811} {\bibfield  {journal} {\bibinfo  {journal} {Proceedings of the Royal Society of London. Series A: Mathematical, Physical and Engineering Sciences}\ }\textbf {\bibinfo {volume} {457}},\ \bibinfo {pages} {1645--1655} (\bibinfo {year} {2001}{\natexlab{b}})}\BibitemShut {NoStop}%
\bibitem [{\citenamefont {Shore}\ and\ \citenamefont {Johnson}(1980)}]{shore-johnson}%
  \BibitemOpen
  \bibfield  {author} {\bibinfo {author} {\bibfnamefont {J.}~\bibnamefont {Shore}}\ and\ \bibinfo {author} {\bibfnamefont {R.}~\bibnamefont {Johnson}},\ }\bibfield  {title} {\enquote {\bibinfo {title} {Axiomatic derivation of the principle of maximum entropy and the principle of minimum cross-entropy},}\ }\href {\doibase 10.1109/TIT.1980.1056144} {\bibfield  {journal} {\bibinfo  {journal} {IEEE Transactions on Information Theory}\ }\textbf {\bibinfo {volume} {26}},\ \bibinfo {pages} {26--37} (\bibinfo {year} {1980})}\BibitemShut {NoStop}%
\bibitem [{\citenamefont {Qian}\ \emph {et~al.}(2009)\citenamefont {Qian}, \citenamefont {Xie},\ and\ \citenamefont {Zhu}}]{qxz-book}%
  \BibitemOpen
  \bibfield  {author} {\bibinfo {author} {\bibfnamefont {M.}~\bibnamefont {Qian}}, \bibinfo {author} {\bibfnamefont {J.-S.}\ \bibnamefont {Xie}}, \ and\ \bibinfo {author} {\bibfnamefont {S.}~\bibnamefont {Zhu}},\ }\href {\doibase 10.1007/978-3-642-01954-8} {\emph {\bibinfo {title} {Smooth Ergodic Theory for Endomorphisms}}}\ (\bibinfo  {publisher} {Springer},\ \bibinfo {address} {Berlin},\ \bibinfo {year} {2009})\BibitemShut {NoStop}%
\bibitem [{\citenamefont {Commons}\ \emph {et~al.}(2021)\citenamefont {Commons}, \citenamefont {Yang},\ and\ \citenamefont {Qian}}]{cyq-21}%
  \BibitemOpen
  \bibfield  {author} {\bibinfo {author} {\bibfnamefont {J.}~\bibnamefont {Commons}}, \bibinfo {author} {\bibfnamefont {Y.-J.}\ \bibnamefont {Yang}}, \ and\ \bibinfo {author} {\bibfnamefont {H.}~\bibnamefont {Qian}},\ }\bibfield  {title} {\enquote {\bibinfo {title} {Duality symmetry, two entropy functions, and an eigenvalue problem in {G}ibbs' theory},}\ }\href {\doibase 10.48550/arXiv.2108.08948} {\bibfield  {journal} {\bibinfo  {journal} {arXiv:2108.08948}\ } (\bibinfo {year} {2021}),\ 10.48550/arXiv.2108.08948}\BibitemShut {NoStop}%
\bibitem [{\citenamefont {Jiang}\ and\ \citenamefont {M.~Qian}(2004)}]{jqq-book}%
  \BibitemOpen
  \bibfield  {author} {\bibinfo {author} {\bibfnamefont {D.-Q.}\ \bibnamefont {Jiang}}\ and\ \bibinfo {author} {\bibfnamefont {M.-P.~Qian}\ \bibnamefont {M.~Qian}},\ }\href {\doibase 10.1007/b94615} {\emph {\bibinfo {title} {Mathematical Theory of Nonequilibrium Steady States: {O}n the Frontier of Probability and Dynamical Systems}}}\ (\bibinfo  {publisher} {Springer-Verlag},\ \bibinfo {address} {Berlin},\ \bibinfo {year} {2004})\BibitemShut {NoStop}%
\bibitem [{\citenamefont {Ge}\ and\ \citenamefont {Qian}(2010)}]{Ge-Qian-2010}%
  \BibitemOpen
  \bibfield  {author} {\bibinfo {author} {\bibfnamefont {H.}~\bibnamefont {Ge}}\ and\ \bibinfo {author} {\bibfnamefont {H.}~\bibnamefont {Qian}},\ }\bibfield  {title} {\enquote {\bibinfo {title} {Physical origins of entropy production, free energy dissipation, and their mathematical representations},}\ }\href {\doibase 10.1103/PhysRevE.81.051133} {\bibfield  {journal} {\bibinfo  {journal} {Phys. Rev. E}\ }\textbf {\bibinfo {volume} {81}},\ \bibinfo {pages} {051133} (\bibinfo {year} {2010})}\BibitemShut {NoStop}%
\bibitem [{\citenamefont {Voigt}(1981)}]{voigt_cmp}%
  \BibitemOpen
  \bibfield  {author} {\bibinfo {author} {\bibfnamefont {J.}~\bibnamefont {Voigt}},\ }\bibfield  {title} {\enquote {\bibinfo {title} {Stochastic operators, information, and entropy},}\ }\href {\doibase 10.1007/BF01941799} {\bibfield  {journal} {\bibinfo  {journal} {Comm. Math. Phys.}\ }\textbf {\bibinfo {volume} {81}},\ \bibinfo {pages} {31--38} (\bibinfo {year} {1981})}\BibitemShut {NoStop}%
\bibitem [{\citenamefont {Rockafellar}(1970)}]{rockafellar-book}%
  \BibitemOpen
  \bibfield  {author} {\bibinfo {author} {\bibfnamefont {R.~T.}\ \bibnamefont {Rockafellar}},\ }\href@noop {} {\emph {\bibinfo {title} {Convex Analysis}}}\ (\bibinfo  {publisher} {Princeton Univ. Press},\ \bibinfo {address} {Princeton},\ \bibinfo {year} {1970})\BibitemShut {NoStop}%
\bibitem [{\citenamefont {Schwinger}(1961)}]{schwinger-61}%
  \BibitemOpen
  \bibfield  {author} {\bibinfo {author} {\bibfnamefont {J.}~\bibnamefont {Schwinger}},\ }\bibfield  {title} {\enquote {\bibinfo {title} {Brownian motion of a quantum oscillator},}\ }\href {\doibase 10.1063/1.1703727} {\bibfield  {journal} {\bibinfo  {journal} {J. Math. Phys.}\ }\textbf {\bibinfo {volume} {2}},\ \bibinfo {pages} {407--432.} (\bibinfo {year} {1961})}\BibitemShut {NoStop}%
\bibitem [{\citenamefont {Keldysh}(1965)}]{keldysh-65}%
  \BibitemOpen
  \bibfield  {author} {\bibinfo {author} {\bibfnamefont {L.~V.}\ \bibnamefont {Keldysh}},\ }\bibfield  {title} {\enquote {\bibinfo {title} {Diagram technique for nonequilibrium processes},}\ }\href@noop {} {\bibfield  {journal} {\bibinfo  {journal} {Sov. Phys. JETP}\ }\textbf {\bibinfo {volume} {20}},\ \bibinfo {pages} {1018--1026} (\bibinfo {year} {1965})}\BibitemShut {NoStop}%
\bibitem [{\citenamefont {Martin}\ \emph {et~al.}(1973)\citenamefont {Martin}, \citenamefont {Siggia},\ and\ \citenamefont {Rose}}]{MSR-73}%
  \BibitemOpen
  \bibfield  {author} {\bibinfo {author} {\bibfnamefont {P.~C.}\ \bibnamefont {Martin}}, \bibinfo {author} {\bibfnamefont {E.~D.}\ \bibnamefont {Siggia}}, \ and\ \bibinfo {author} {\bibfnamefont {H.~A.}\ \bibnamefont {Rose}},\ }\bibfield  {title} {\enquote {\bibinfo {title} {Statistical dynamics of classical systems},}\ }\href {\doibase 10.1103/PhysRevA.8.423} {\bibfield  {journal} {\bibinfo  {journal} {Phys. Rev. A}\ }\textbf {\bibinfo {volume} {8}},\ \bibinfo {pages} {423--437} (\bibinfo {year} {1973})}\BibitemShut {NoStop}%
\bibitem [{\citenamefont {Parrondo}\ \emph {et~al.}(2015)\citenamefont {Parrondo}, \citenamefont {Horowitz},\ and\ \citenamefont {Sagawa}}]{thermo_info}%
  \BibitemOpen
  \bibfield  {author} {\bibinfo {author} {\bibfnamefont {J.}~\bibnamefont {Parrondo}}, \bibinfo {author} {\bibfnamefont {J.}~\bibnamefont {Horowitz}}, \ and\ \bibinfo {author} {\bibfnamefont {T.}~\bibnamefont {Sagawa}},\ }\bibfield  {title} {\enquote {\bibinfo {title} {Thermodynamics of information},}\ }\href {\doibase 10.1038/nphys3230} {\bibfield  {journal} {\bibinfo  {journal} {Nature Phys.}\ }\textbf {\bibinfo {volume} {11}},\ \bibinfo {pages} {131--139} (\bibinfo {year} {2015})}\BibitemShut {NoStop}%
\bibitem [{\citenamefont {Ge}\ and\ \citenamefont {Qian}(2012)}]{ge_qian_ijmpb}%
  \BibitemOpen
  \bibfield  {author} {\bibinfo {author} {\bibfnamefont {H.}~\bibnamefont {Ge}}\ and\ \bibinfo {author} {\bibfnamefont {H.}~\bibnamefont {Qian}},\ }\bibfield  {title} {\enquote {\bibinfo {title} {Analytical mechanics in stochastic dynamics: Most probable path, large-deviation rate function and {H}amilton–{J}acobi equation},}\ }\href {\doibase 10.1142/S0217979212300125} {\bibfield  {journal} {\bibinfo  {journal} {International Journal of Modern Physics B}\ }\textbf {\bibinfo {volume} {26}},\ \bibinfo {pages} {1230012} (\bibinfo {year} {2012})}\BibitemShut {NoStop}%
\bibitem [{\citenamefont {Qian}\ \emph {et~al.}(2002)\citenamefont {Qian}, \citenamefont {Qian},\ and\ \citenamefont {Tang}}]{qqt_jsp}%
  \BibitemOpen
  \bibfield  {author} {\bibinfo {author} {\bibfnamefont {H.}~\bibnamefont {Qian}}, \bibinfo {author} {\bibfnamefont {M.}~\bibnamefont {Qian}}, \ and\ \bibinfo {author} {\bibfnamefont {X.}~\bibnamefont {Tang}},\ }\bibfield  {title} {\enquote {\bibinfo {title} {Thermodynamics of the general diffusion process: {T}ime-reversibility and entropy production},}\ }\href {\doibase 10.1023/A:1015109708454} {\bibfield  {journal} {\bibinfo  {journal} {J. Stat. Phys.}\ }\textbf {\bibinfo {volume} {107}},\ \bibinfo {pages} {1129--1141} (\bibinfo {year} {2002})}\BibitemShut {NoStop}%
\bibitem [{\citenamefont {Fowler}(1929)}]{fowler-book}%
  \BibitemOpen
  \bibfield  {author} {\bibinfo {author} {\bibfnamefont {R.~H.}\ \bibnamefont {Fowler}},\ }\href@noop {} {\emph {\bibinfo {title} {Statistical Mechanics: The Theory of the Properties of Matter in Equilibrium}}}\ (\bibinfo  {publisher} {Cambridge Univ. Press},\ \bibinfo {address} {London},\ \bibinfo {year} {1929})\BibitemShut {NoStop}%
\bibitem [{\citenamefont {Goldenfeld}(1992)}]{landau}%
  \BibitemOpen
  \bibfield  {author} {\bibinfo {author} {\bibfnamefont {N.}~\bibnamefont {Goldenfeld}},\ }\href@noop {} {\emph {\bibinfo {title} {Lectures on Phase Transitions and the Renormalization Group}}}\ (\bibinfo  {publisher} {Addison-Wesley},\ \bibinfo {address} {Reading, MA},\ \bibinfo {year} {1992})\BibitemShut {NoStop}%
\end{thebibliography}

\appendix

\section{Legendre-Fenchel and Laplace/Fourier Transforms}
\label{appendix}

There is an underlying mathematical connection between the Legendre-Fenchel transform (LFT) and the Laplace/Fourier-type integral transform that is employed in the computation of partition functions in Gibbs' theory.  The latter reduces to the former in the asymptotic limit of large number, represented by the $\epsilon\to 0$ below, as in the Laplace's method for asymptotic evaluation of integrals.  Known as the Darwin-Fowler method \cite{fowler-book}, this is the mathematical basis for the derivation, in the thermodynamic limit, of the relationship between thermodynamic potentials connected through the LFT from the statistical equilibrium ensemble change where different partition functions are computed via integral transforms.

The last step in Eq. (\ref{eq41}) is due to the following reasoning: Since we search $\vx'$ for the infimum, any $\vx'$ that results $\{\cdots\} =\infty$ is negligible. This is the case for all $\vx'$ with $\mT^*\vx'\neq\vx$, when the supremum of the linear function $(\vx-\mT^*\vx')\cdot\vy$ is precisely $\infty$ for $\vy\in\mathbb{R}^K$.
Therefore the only relevant $\vx'$ is restricted by $\mT^*\vx'=\vx$.

We now derive Eq. (\ref{eq41}) through the integral transform.  Corresponding to the LFT between the pair of functions $\tilde{\phi}(\vx)$ and $\tilde{\psi}(\vy)$ in Eq. (\ref{eq37}), the integral transform reads
\begin{eqnarray}
    e^{\frac{\tilde{\phi}(\vx)}{\epsilon}} &=& \int_{\mathbb{R}^K} \rd\vy\ e^{\frac{\vx\cdot\vy - \tilde{\psi}(\vy)}{\epsilon} }
    = \int_{\mathbb{R}^K}\rd\vy\ e^{\frac{ \vx\cdot\vy - \psi(\mT\vy)}{\epsilon} }
\nonumber\\
   &=& \int_{\mathbb{R}^K} \rd\vy\ e^{\frac{\vx\cdot\vy}{\epsilon}}\int_{\mathbb{R}^K} \rd\vx'\ e^{-\frac{\vx'\cdot\mT\vy-\phi(\vx')}{\epsilon}}
\nonumber\\
   &=& \int_{\mathbb{R}^K} \rd\vx'\ e^{\frac{\phi(\vx')}{\epsilon}}\int_{\mathbb{R}^K} \rd\vy\ e^{-\frac{(\mT^{*}\vx')\cdot\vy-\vx\cdot\vy}{\epsilon}}
\nonumber\\
   &=& \int_{\mathbb{R}^K} \rd\vx'\ e^{\frac{\phi(\vx')}{\epsilon}} \delta\big(\mT^{*}\vx'-\vx\big),
\label{eq42}
\end{eqnarray}
which yields Eq. (\ref{eq41}) in the asymptotic limit of $\epsilon\to 0$.

From Eq. (\ref{eq42}), one also recognizes that
\begin{eqnarray}
\tilde{\phi}(\vx) &=& \epsilon\log \int_{\mathbb{R}^K} \rd\vx'\ e^{\frac{\phi(\vx')}{\epsilon} } \delta\big(\mT^{*}\vx'-\vx\big),
\end{eqnarray}
which is exactly the negative of Landau's effective Hamiltonian/free energy defined on the state space $\vx$ obtained through a coarse-graining operation over the original state space $\vx'$ under the constraint specified by $\delta(\mT^{*}\vx'-\vx)$ \cite{landau}. The power of Eqs. (\ref{eq41}) or (\ref{eq42}) lies in that the effective Hamiltonian can be directly obtained through a LFT of the cumulant generating function $\psi(\mT\vy)$ as in (\ref{eq41}), or equivalently through an integral transform as in (\ref{eq42}) which reduces to LFT in the asymptotic limit:
\begin{eqnarray}
\tilde{\phi}(\vx) &=& \lim_{\epsilon\to 0}\ \epsilon\log \int_{\mathbb{R}^K} \rd\vy\ e^{\frac{ \vx\cdot\vy - \psi(\mT\vy)}{\epsilon} } = \sup_{\vy\in\mathbb{R}^K}\Big\{\vx\cdot\vy
 -\psi(\mT\vy) \Big\}.
\end{eqnarray}
Equations as such establish the equivalence between Gibbs' statistical ensemble theory in $\epsilon^{-1}\to\infty$ limit and thermodynamics.
\end{document}